\documentclass[twocolumn,superscriptaddress,floatfix,preprintnumbers]{revtex4-1}

\usepackage[ colorlinks = true,
             linkcolor = blue,
             urlcolor  = blue,
             citecolor = blue,
             anchorcolor = green,
]{hyperref}

\usepackage{amsmath}
\usepackage{amssymb}
\usepackage{graphicx}
\usepackage{epsfig}
\usepackage{braket}
\usepackage{slashed}
\usepackage{bm}
\usepackage[usenames,dvipsnames]{color}
\usepackage{color}

\usepackage[caption=false]{subfig}

\usepackage{mathtools}
\mathtoolsset{showonlyrefs}

\newcommand{\tr}{\mathtt{r}}

\def\idty{{\mathbb{I}}}

\def\bra #1{\langle #1\vert}
\def\ket #1{\vert #1\rangle}

\def\ketbra #1#2{\vert #1\rangle \langle #2\vert}
\def\kettbra#1{\ketbra{#1}{#1}}
\def\tr{{\rm Tr}}
\newcommand\Scp[2]{\ensuremath{\, \langle #1 \,\vert #2 \,\rangle}}

 \newcommand*{\cA}{\mathcal{A}}
 \newcommand*{\cB}{\mathcal{B}}
 \newcommand*{\cC}{\mathcal{C}}

 \newcommand*{\cG}{\mathcal{G}}
 \newcommand*{\cH}{\mathcal{H}}

 \newcommand*{\cQ}{\mathcal{Q}}

\begin{document}

\title{The entanglement of distillation for gauge theories
} 

\author{Karel Van Acoleyen}
\affiliation{Department of Physics and Astronomy, Ghent University, Krijgslaan 281, S9, 9000 Gent, Belgium}

\author{Nick Bultinck}
\affiliation{Department of Physics and Astronomy, Ghent University, Krijgslaan 281, S9, 9000 Gent, Belgium}
\author{Jutho Haegeman}
\affiliation{Department of Physics and Astronomy, Ghent University, Krijgslaan 281, S9, 9000 Gent, Belgium}
\author{Michael Marien}
\affiliation{Department of Physics and Astronomy, Ghent University, Krijgslaan 281, S9, 9000 Gent, Belgium}
\author{Volkher B.~Scholz}
\affiliation{Department of Physics and Astronomy, Ghent University, Krijgslaan 281, S9, 9000 Gent, Belgium}
\author{Frank Verstraete}
\affiliation{Department of Physics and Astronomy, Ghent University, Krijgslaan 281, S9, 9000 Gent, Belgium}
\affiliation{Vienna Center for Quantum Science and Technology, Faculty of Physics, University of Vienna, Boltzmanngasse 5, 1090 Vienna, Austria}

\begin{abstract}

  We study the entanglement structure of lattice gauge theories from the local operational point of view, and, similar to Soni and Trivedi (arXiv:1510.07455), we show that the usual entanglement entropy for a spatial bipartition can be written as the sum of an undistillable gauge part and of another part corresponding to the LOCC distillable entanglement, which is obtained by depolarizing the local superselection sectors. We demonstrate that the distillable entanglement is zero for pure abelian gauge theories in the weak coupling limit, while it is in general nonzero for the nonabelian case. We also consider gauge theories with matter, and show in a perturbative approach how area laws -- including a topological correction -- emerge for the distillable entanglement.
%
\end{abstract}

\maketitle
%
%
%
%
\noindent\emph{Introduction}---%
The concept of quantum entanglement is at the center of current day physics. While it used to be confined to the realms of quantum information theory, it is now appearing in many other fields. It is used for instance to characterize new phases of matter with topological order~\cite{PhysRevLett.96.110404,PhysRevLett.96.110405}, and it also plays a crucial role as guiding principle in the classical simulation of general quantum many body systems~\cite{RevModPhys.82.277,reviewMPSPEPS,MERAintro}. Furthermore in the context of the holographic principle and quantum gravity there appears to be an intricate connection between entanglement and geometry \cite{2010GReGr..42.2323V,2006PhRvL..96r1602R}. At the same time entanglement lies at the root of the firewall paradox in the quantum behavior of black holes \cite{firewall,braunstein}.

It is hence highly desirable to gain a better understanding of the entanglement structure of quantum field theories. However, as first pointed out in \cite{Buividovich:2008jp}, for gauge theories the concept of entanglement is obscured by their intrinsic non-locality. More formally, the Hilbert space for locally distinct regions is not of tensor product form, which renders the usual rules for computing the entanglement not applicable. While a number of tools have been developed to cope with this computational problem~\cite{Buividovich:2008jp,Donnely,Casini,Aoki:2015gg,Gromov_2014,2015arXiv150908478R,2014arXiv1404.1391R,Ghosh:2015aa,Hung:2015aa}, the physical meaning of entanglement in gauge theories was still unclear.

In this letter, we put forward the operational view on entanglement: for us, the physical entanglement is identified as the \emph{accessible} one. That is, the asymptotic number of EPR-pairs $E_D^{\text{gauge}}$ which can be distilled \cite{PhysRevA.53.2046,wilde2013quantum} per copy from many copies of the lattice gauge theory state by physically allowed local operations in combination with classical communication (LOCC) acting on spatially separated regions of the lattice. We show that from this operational point of view, every pure state is physically undistinguishable from a mixed state if we are only allowed to use local gauge preserving operations. Hence it is the distillable entanglement of this mixed state which is physically relevant. The expression we find for the distillable entanglement in a lattice gauge theory was already identified as being part of the overall mathematical expression for the entanglement entropy in earlier works~\cite{Buividovich:2008jp,Donnely,Casini,Aoki:2015gg,Gromov_2014,2015arXiv150908478R,2014arXiv1404.1391R,Ghosh:2015aa,Hung:2015aa}, however, its operational interpretation was missing so far (but see also the very recent work~\cite{2015arXiv151007455S}, where similar conclusions were reached). Moreover, our results imply that it is the \emph{only} physical sensible definition of entanglement, as it is defined with respect to the physical observables. As such, it adds to recent progresses in understanding the entanglement structure of physical systems obeying superselection rules in operational terms~\cite{VollbrechtWernerSymmetries,PhysRevA.90.062325,PhysRevLett.91.010404,PhysRevLett.92.087904} and can be seen as another instance of the ``think operationally'' paradigm approach to quantum physics~\cite{PhysRevLett.92.107902}.

%
%
We illustrate our findings by calculating the physically relevant entanglement for ground states of lattice gauge theories in the weak coupling limit, both for pure gauge theories and for theories involving matter.
%
Furthermore we show in a perturbative approach how an area law with a topological correction emerges for the distillable entanglement of  $\mathbb{Z}_2$ gauge theory with matter and argue that the topological correction is generic for zero-coupling abelian gauge theories with finite groups.
%

\begin{figure}[t]
  \subfloat[]{\label{fig:1}}{\includegraphics[scale=0.55]{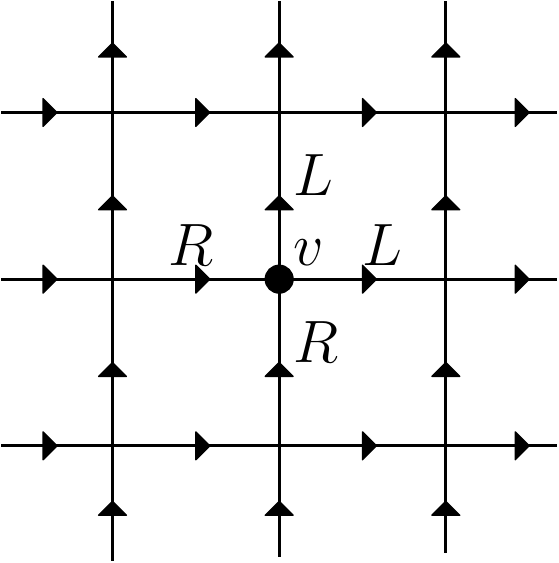}} \qquad
  \subfloat[]{\label{fig:2}}{\includegraphics[scale=0.55]{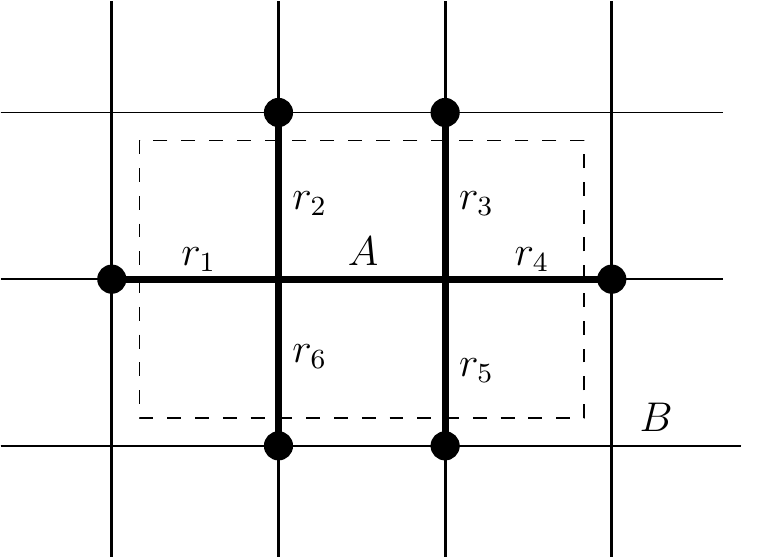}}
  \caption{(a) The setting of lattice gauge theory: the action of a gauge transformation $U_v$ is determined by the orientation of the edges. (b) A bipartite cut: the edges along the cut are labeled by irreducible representations $r_1,\ldots,r_6$ of the group action at the 6 different boundary vertices (thick points), giving rise to a direct sum structure. Fixing theses irreducible representations produces a direct product structure on each boundary edge, with one factor lying on the inside and one factor on the outside.}
\end{figure}

\noindent\emph{Hamiltonian lattice gauge theory}---%
We only briefly review the setup of the Kogut-Susskind (K-S) Hamiltonian formulation of lattice gauge theory~\cite{PhysRevD.11.395,Kogut:1979jp,BaezLGT} in order to extract the relevant facts for us. A more detailed discussion can be found in the Appendix or in the literature~\cite{PhysRevD.11.395,Kogut:1979jp,BaezLGT,Casini,Donnely}. We start with a compact group $G$, as well as with a lattice $(V,L)$ consisting of vertices $v \in V$ that are connected by edges $e \in L$. The gauge degrees of freedom are located on the edges: on each edge $e$ we have a local Hilbert space $\mathcal{H}_e \simeq L^2(G)$ on which the group acts by left ($L_e$) and right ($R_e$) multiplication, (Fig.~\ref{fig:1}). The matter degrees of freedom are located on the vertices: on each vertex $v$ the local Hilbert space $\mathcal{H}_m$ consists of a representation space of the group $G$, with a certain number of irreducible representations and the local group action $V_v$ defined accordingly. The overall Hilbert space $\cH$ is given by the tensor product of all local ones. In order to define the local gauge transformations, we assign an orientation to each  edge. The local gauge transformation at vertex $v$ is given by the operator $U_v(g)=V_v(g)\otimes_{e\in E^+_v} R_e(g)\otimes_{e \in E^-_v}L_{e}(g)$, where $E^+_v$ (resp. $E^-_v$) denote the sets of all incoming (resp. outgoing) edges of the vertex $v$. The subspace $\mathcal{H}_{\text{phys}}$ of physical states $\ket{\psi}$ is singled out by the constraint $U_v(g)\ket{\psi} = \ket{\psi}$ for all $v \in V$ and $g \in G$. Correspondingly, the algebra $\mathcal{B}(\mathcal{H}_{phys})$ of physically allowed operations $O$, consist of all gauge invariant operators on $\mathcal{H}$, that is those operators that obey $ U^+_v(g)O U_v(g)=O$, again for all $v$ and $g$. The non-local character of gauge theories is now manifest: since $U_v(g)$ is not 1-local, having non-trivial support on a vertex $v$ and different edges $e$, the true physical Hilbert space $\mathcal{H}_{\text{phys}}$ does not decompose in a direct product of local Hilbert spaces. As such, the tensor product structure of the Hilbert space $\mathcal{H}$ is nothing but a convenient illusion that is used to define the true ``non-local'' Hilbert space $\mathcal{H}_{\text{phys}}$. But notice that we still have a clear notion of local observables and operations: to each gauge invariant operator $O$ we can assign the spatial region of non-trivial support.

\noindent\emph{Locality and allowed operations}---%
Let us make the non-locality more concrete by considering some bipartition of the full system in a connected spatial region $A$ and its complement $B$, connected by $n$ boundary edges (Fig.~\ref{fig:2}). The structure of the associated local Hilbert spaces $\cH_A$, $\cH_B$ is most apparent if written in the representation bases corresponding to the gauge transformations crossing the boundary (see the Appendix for more details). Specifically for $A$ we define the basis $\{\ket{\vec{r},\vec{i},\alpha}_A\}$, with $\vec{r}=(r_1,r_2,\ldots,r_n) \in \Lambda^n$, where $\Lambda$ is the set of equivalence classes of irreducible representations of $G$, with the subindex running over the $n$ different boundary lattice vertices $v_b$, and $\vec{i}=(i_1,i_2,\dots, i_n)$ enumerating a basis in the associated representation space. For each $\vec{r}$ the multiplicity space is labelled by  $\alpha$ (which has zero range if $\vec{r}$ does not appear in the basis). In a similar way we can take for $\mathcal{H}_B$ the irreducible representation basis of $\otimes_{v_b} G$ on $B$: $\{\ket{\vec{r},\vec{i},\beta}_B\}$, with the difference that we now label the states according to their transformation under the complex conjugate irreducible representations. The local Hilbert spaces thus have the following direct sum structure:\begin{align}
   \mathcal{H}_A = \oplus_{\vec{r}} \mathcal{H}^{\vec{r}}_{Ag} \otimes \mathcal{H}^{\vec{r}}_{Am}\,,\quad\quad \mathcal{H}_B = \oplus_{\vec{r}} \mathcal{H}^{\vec{r}}_{Bg} \otimes \mathcal{H}^{\vec{r}}_{Bm}\,,
\label{eq:decompAB} \end{align}
with $\mathcal{H}^{\vec{r}}_{Ag}, \mathcal{H}^{\vec{r}}_{Bg}$ the representation spaces for the irrep $\vec{r}$ corresponding to the group indices $\vec{i}$ and $\mathcal{H}^{\vec{r}}_{Am}, \mathcal{H}^{\vec{r}}_{Bm}$ the multiplicity spaces on $A$ and $B$.

Now, the gauge constraints imply that all physical states have to be maximally entangled between the representation spaces, which amounts to the decomposition
\begin{align}
  \label{eq:decompphyshilbertspace}
  \mathcal{H}_{\text{phys}}= \bigoplus_{\vec{r} \in \Lambda^n} \left(\ket{\phi^{\vec{r}}}_{g AB}\otimes \mathcal{H}^{\vec{r}}_{Am} \otimes \mathcal{H}^{\vec{r}}_{Bm}\right) \,,
\end{align}
with $\ket{\phi^{\vec{r}}}_{g AB}=\sum_{\vec{i}}\ket{\vec{i}}_A\ket{\vec{i}}_B/\sqrt{d_{\vec{r}}}$, the maximally entangled state in the representation space $\mathcal{H}^{\vec{r}}_{Ag} \otimes \mathcal{H}^{\vec{r}}_{Bg}$. Here $d_{\vec{r}}=\prod_{v_b}d_{r_{v_b}}$ denotes the dimension of the direct product irrep $\vec{r}$. Notice that for the $A/B$ partition the `inner' and `outer' gauge constraints, for which the gauge transformations $U_v(g)$ only have non-trivial support on either $A$ or $B$, can be considered local. The effect of these constraints is taken into account implicitly by restricting the multiplicity spaces accordingly.  %

It is now evident that indeed $\mathcal{H}_{\text{phys}}\neq\mathcal{H}_A\otimes\mathcal{H}_B$. However, after projecting onto a specific choice of boundary representations $\vec{r}$, the physical Hilbert space is isomorphic to the well-defined tensor product Hilbert space $\mathcal{H}^{\vec{r}}_{Am} \otimes \mathcal{H}^{\vec{r}}_{Bm}$ separating the region $A$ from its complement $B$. 
Since physical operations which are either located in the inside or in the outside of the region $A$ cannot change the superselection sector labeled by $\vec{r}$, they have to commute with the projection $P_{\vec{r}} = \kettbra{\phi^{\vec{r}}}_{g AB}\otimes \idty_{\mathcal{H}^{\vec{r}}_{Am}} \otimes \idty_{\mathcal{H}^{\vec{r}}_{Bm}} = W_{\vec{r}} W_{\vec{r}}^\dagger$ where the isometry $W_{\vec{r}}=\ket{\phi^{\vec{r}}}_{g AB}\otimes \idty_{\mathcal{H}^{\vec{r}}_{Am}} \otimes \idty_{\mathcal{H}^{\vec{r}}_{Bm}} $ satisfies $W_{\vec{r}}^\dagger\cH_{\text{phys}} \simeq \mathcal{H}^{\vec{r}}_{Am} \otimes \mathcal{H}^{\vec{r}}_{Bm}$.
%
%
%
%
From an algebraic viewpoint, this implies that the algebra of physical local observables has a non-trivial center, as already pointed out in~\cite{Casini}. We find that the algebra $\cC$ of local physical observables for the $A/B$ partition equals
%
\begin{align}
  \label{eq:physicalcutobservables}
  \cC = \bigoplus_{\vec{r} \in \Lambda^{n}} \cA_A(\vec{r}) \otimes \cA_A(\vec{r})^\prime \subset \cB(\cH_{\text{phys}})\,,
\end{align}
where $\cA_A(\vec{r}) = P_{\vec{r}} \cA_A P_{\vec{r}}$ is the projected algebra $\cA_A$ of all gauge invariant operators acting on the region $A$ and $\cA_A(\vec{r})^\prime$ is the commutant of $\cA_A(\vec{r})$, i.e. all operators in $\cB(P_{\vec{r}}\cH_{\text{phys}})$ which commute with $\cA_A(\vec{r})$.
%
The role of the commutation relation is to ensure locality: as all elements in $\cA_A(\vec{r})^\prime$ commute with elements in $\cA_A(\vec{r})$ and map physical states to physical states 
they correspond to the operations which can be performed by acting only on the outside of the region $A$.

%
%
\noindent\emph{Equivalence of pure and mixed states}---%
The structure of gauge theories imply that local operations $O \in \cC$ are not only constrained by geometry but also by the gauge constraints. Specifically, exploiting the fact that they commute with the orthogonal projections $P_{\vec{r}}$ as well as using 
%
$\sum_{\vec{r}} P_{\vec{r}}\ket{\psi} = \ket{\psi}$ for $\ket{\psi} \in \cH_{\text{phys}}$, we find for the expectation values for elements $O \in \cC$ that
\begin{align}
  \label{eq:}
  \Scp{\psi}{O \psi} &= 
  \sum_{\vec{r}\in \Lambda^{n}} \tr[O P_{\vec{r}} \kettbra{\psi}P_{\vec{r}}] \,.
\end{align}
Therefore physical operations respecting the bipartite cut cannot distinguish between the pure state $\kettbra{\psi}$ and the mixed depolarized state
\begin{align}
  \label{eq:depolarizedstate}
  \sigma^\psi = \sum_{\vec{r}\in \Lambda^{n}} P_{\vec{r}} \kettbra{\psi}P_{\vec{r}}\,.
\end{align}
This implies that from the point of view of a physical experiment, the entanglement of the pure state $\psi$ is irrelevant and only the mixed state $\sigma^\psi$ is physical. We will show that the physically accessible entanglement is indeed the one contained in the latter. That symmetry constraints can lead to such an equivalence of pure and mixed states was first noted in~\cite{PhysRevLett.92.107902}.

%
\noindent\emph{Main result}---%
As first suggested in \cite{Buividovich:2008jp}, to define a formal bipartite entanglement entropy for an arbitrary physical state $\ket{\psi}\in \mathcal{H}_{\text{phys}}$, one can consider it as a state on the larger local Hilbert space $\mathcal{H}=\mathcal{H}_A\otimes \mathcal{H}_B$
and trace out the subsystem $B$ to get a density matrix $\rho_A=Tr_B\ket{\psi}\bra{\psi}$. As was first shown in \cite{Donnely} for the general nonabelian case, the corresponding von Neumann entropy can then be decomposed as:
\begin{align}
  \label{eq:fullentropy}
  S_A = -\tr_{\!A}\rho_A\ln\rho_A = -\sum_{\vec{r}}p_{\vec{r}}\ln \frac{p_{\vec{r}}}{d_{\vec{r}}}+\sum_{\vec{r}}p_{\vec{r}}S_A^{\vec{r}}
\end{align}
with $S_A^{\vec{r}}=-\tr\rho_A^{\vec{r}}\ln \rho_A^{\vec{r}}$ the von Neumann entropy of the density matrix on the multiplicity space of the irrep sector $\vec{r}$, $\rho_A^{\vec{r}} = \frac{1}{p_{\vec{r}}} \tr_{\mathcal{H}^{\vec{r}}_{Bm}}[W_{\vec{r}}^\dagger \kettbra{\psi}W_{\vec{r}}]$, the normalization given by $p_{\vec{r}} = \Scp{\psi}{P_{\vec{r}}|\psi}$.
However, as discussed before, the pure state $\ket{\psi}$ is physically equivalent to the depolarized mixed state $\sigma^\psi$ and hence for operational purposes such as entanglement distillation, this state has to be considered. As shown below, the fact that the eigenstates of $\sigma^\psi$ can be locally distinguished by acting only on the inside or outside of the region $A$ leads to the following expression of the entanglement of distillation for $\sigma^\psi$,
\begin{align}
  \label{eq:mainresult}
  E_D^{\text{gauge}}(\ket{\psi}) = E_D(\sigma^\psi) = \sum_{\vec{r}}p_{\vec{r}}S_A^{\vec{r}}\,,
\end{align}
which equals the second term in Eq.~\eqref{eq:fullentropy} (see also \cite{2015arXiv151007455S}). Hence it is this term which possesses a clear
operational meaning of the true physical quantum entanglement; and in contrast to previous discussions in the literature we consider the first term to be an artefact of the embedding of the physical Hilbert space~$\mathcal{H}_{\text{phys}}$ into the tensor product space~$\mathcal{H}$.

%
\noindent\emph{Proof}---%
  The proof is naturally divided into two parts: the statement that this amount of entanglement \emph{can} be extracted (direct part), and the converse statement that \emph{no more} entanglement can be extracted.
  In order to proceed, we first remark that the projected algebra  $\cA_A(\vec{r})$ together with its commutant
  spans the algebra of all operators on the projected subspace $P_{\vec{r}}\cH_{\text{phys}}$,
    which is equivalent to the already established fact that this subspace decomposes into a tensor product of two Hilbert spaces. This implies that after projecting (and adjusting the normalization) the state $\ket{\psi}$ onto the subspace $P_{\vec{r}}\cH_{\text{phys}}$, we are again in the well-known situation of standard quantum information theory: we are given a pure state on a tensor product, and all operations on each tensor factor are allowed local operations. In this case the formula of the entanglement of distillation is given by the entropic expression $S_A^{\vec{r}}$ which also equals the entanglement of formation (that is, the entanglement necessary to asymptotically create the state by local operations and classical communication).
  The direct part then proceeds as follows: Each party first measures with respect to the set of projectors $\{P_{\vec{r}}\}_{\vec{r} \in \Lambda^{n}}$. On average, this brings the state into the form of the depolarized state $\sigma^\psi$ (cf. Eq.~\eqref{eq:depolarizedstate}). Depending on the measurement outcome, each one appearing with probability $p_{\vec{r}}$, they are left with the pure state $\frac{1}{p_{\vec{r}}} P_{\vec{r}} \kettbra{\psi}P_{\vec{r}}$. They then perform the usual entanglement distillation protocol~\cite{PhysRevA.53.2046,wilde2013quantum}, asymptotically yielding $S_A^{\vec{r}}$ amount of entanglement. On average of all outcomes, this gives an asymptotic rate of $\sum_{\vec{r} \in \Lambda^{n}} p_{\vec{r}} S_A^{\vec{r}}$ many ebits.
 For the converse part, it suffices to show that the entanglement cost $E_C^{\text{gauge}}$ of $\sigma^\psi$ --- that is, the asymptotic rate of ebits needed to successfully create many copies of the state by LOCC, a task called \emph{entanglement dilution}~\cite{PhysRevA.53.2046,wilde2013quantum,Hayden_2001} --- equals $\sum_{\vec{r} \in \Lambda^{n}} p_{\vec{r}} S_A^{\vec{r}}$. This follows from $E^{\text{gauge}}_D\leq E^{\text{gauge}}_C$, since otherwise the two parties could interchangeably run the two protocols to create entanglement by LOCC, which is absurd. We start with one party creating the overall state $\kettbra{\psi}$ locally on one side. We proceed to measure with respect to the projections onto boundary representations, sending the value of $\vec{r}$, and conditioned on this measurement result applying the usual entanglement dilution protocol~\cite{PhysRevA.53.2046,wilde2013quantum,Hayden_2001} for the state $\frac{1}{p_{\vec{r}}} P_{\vec{r}} \kettbra{
 \psi}P_{
 \vec{r}}$, asymptotically using up $S_A^{\vec{r}}$ pairs of ebits. On average, this protocol creates the state $\sigma^\psi$ while using $\sum_{\vec{r} \in \Lambda^{n}} p_{\vec{r}} S_A^{\vec{r}}$ many ebits.

 As a side result, we find that both operationally defined measures of entanglement, the distillable entanglement $E_D^{\text{gauge}}$ and the entanglement cost $E_C^{\text{gauge}}$ are equal to each other and given by~\eqref{eq:mainresult} if we restrict to local gauge invariant operations and classical communication.


%

\noindent\emph{Weak coupling results for pure lattice gauge theories}---%
For a pure lattice gauge theory, the zero-coupling Hamiltonian is given as the sum over all Wilson loops around plaquettes, $ H_0 = -\sum_{\textrm{Plaquettes } p} W_p$,
where for a plaquette $p$ with edges $e=1,2,3,4$, $W_p$ is a projector diagonal in the group basis, projecting onto the states $\ket{g_1,g_2,g_3,g_4}$ for which $\prod_i g_i^{p_i}=e$. Here $e$ stands for the group identity element and $p_i=+1$ for a clockwise orientation of the edge $i$ along the plaquette and $p_i=-1$ for a counterclockwise orientation. As $W_p$ is gauge invariant, the zero-coupling ground state $\ket{\psi_0}$ is given by applying the projector $\cQ_{\cH_{\text{phys}}}=\sum_{\vec{r}}P_{\vec{r}}$ on the product state with all edge states initialized to the group identity.

For abelian groups the ground state of $H_0$ has no distillable entanglement. For concreteness, consider a $\mathbb{Z}_2 = \{e,x\}$ zero-coupling gauge theory, also called the Toric code \cite{kitaev03}, where we take $R(x) = L(x) = \sigma_x$. The corresponding plaquette term $W_p$ is given by $1/2(\mathbb{I}^{\otimes 4} + \sigma_z^{\otimes 4})$. The ground state of $H_0$ is then simply the equal weight superposition of all closed loop configurations of $|-\rangle$ states on the edges, where $\sigma_x|-\rangle = -|-\rangle$. The projectors $P_{\vec{r}}$ now simply project the boundary edges into a particular configuration of $|+\rangle$ and $|-\rangle$ states. The resulting state $W_{\vec{r}} \ket{\psi_0}$ is easily seen to be a product state $\ket{\phi_{Am}}\otimes\ket{\phi_{Bm}}$, from which we conclude that the distillable entanglement is zero. This argument is readily generalized to arbitrary abelian groups (and even to twisted versions of $H_0$ \cite{DijkgraafWitten,twistedQD} such as the double semion model).

We remark that abelian theories already show that the distillable entanglement does not satisfy strong subadditivity~\cite{SSA,2015arXiv151102671M}, and even fails to fulfill subadditivity: A simple example is a state for a $\mathbb{Z}_2$-abelian gauge theory which is in a superposition of having two closed loops, one crossing the boundary between two regions $A$ and $B$ and one outside of both, and no loops. Operations either located inside $A$ or $B$ cannot detect the loop crossing the boundary --- which implies that the entanglement distillable between the regions $A$ or $B$ and their complements is the same as for the states with no loops present: zero. On the contrary, operations located in the union $A \cup B$ \emph{can} detect the loop present and hence can exploit the entanglement between the loops inside and outside of $A \cup B$. This example can be amplified to yield a violation of subadditivity scaling with the area of the boundary between $A$ and $B$. The violation can be traced back to the fact that for every choice of a bipartite cut, the center of the observable algebra chances, implying a different depolarization map for different cuts.

Let us now calculate the distillable entanglement in the ground state for nonabelian theories at zero coupling. 
For these, 
we find that the reduced density matrix has flat spectrum with respect to representation basis $\ket{\vec{r},\vec{i},\alpha}$, and for fixed $\vec{r}$ the rank is given by the number $N^1_{\vec{r}}$ of inequivalent ways the representations $\vec{r}$ can fuse to the scalar one (see the Appendix for details).
%
%
%
%
This gives a distillable entanglement \eqref{eq:mainresult} of
\begin{align}
  \label{eq:nonabelian}
    E_D^{\text{gauge}}(\ket{\psi_0}) =\frac{1}{|G|^{n-1}}\sum_{\vec{r} \in \Lambda^n} d_{\vec{r}}N^1_{\vec{r}}\log N^1_{\vec{r}}\,.
\end{align}
For abelian groups $N_{\vec{r}}=0,1$ and we find $E_D^{\text{gauge}}(\ket{\psi_0})=0$ consistent with the derivation above, for nonabelian groups $N_{\vec{r}}\neq 0,1$ for some $\vec{r}$, implying a nonzero distillable entanglement already at zero coupling.

%
%
\noindent\emph{Theories with matter}--- Let us now consider the distillable entanglement of a globally symmetric matter state $|\psi_{\text{mat}}\rangle$ after gauging. For simplicity we restrict to abelian groups. Before gauging, the entanglement entropy of the matter state after tracing out $B$ is given by $S(\rho_A) = H(p_r) + \sum_r p_r S(\rho^A_r)$, where $p_r = \langle\psi_{\text{mat}}|\Pi^A_r| \psi_{\text{mat}}\rangle$ is the probability of measuring total charge $r$ in $A$, $H(p_r)$ is the Shannon entropy of $\{p_{r}\}$,  and $p_r\rho^A_r = \tr_B\left(\Pi^A_r|\psi_{\text{mat}}\rangle\langle \psi_{\text{mat}}|\Pi^A_r \right)$. Analogous to the case of local gauge constraints, the restriction to symmetric operations implies that the state is again equivalent to a depolarized mixed state $\sigma^{\psi_{\text{mat}}}$, and the  entanglement distillable by symmetric operations of the matter state is given by the second term $\sum_r p_r S(\rho^A_r)$. For this argument to hold it is important that when we consider multiple copies in order to implement the distillation protocol, we demand that the symmetry conditions hold for each copy simultaneously, not only just for all copies at once, because otherwise also the first term is distillable~\cite{Schuch_2008}.

 We can now apply the procedure of Ref.~\cite{gaugingpaper} to obtain the gauged state $|\psi_{g,\text{mat}}\rangle$ in the zero coupling limit: $|\psi_{g,\text{mat}}\rangle = \mathcal{G}|\psi_{\text{mat}}\rangle$, with $\mathcal{G}$ an isometry called the \emph{gauging map} that maps to the Hilbert space containing both matter and gauge degrees of freedom on the vertices and edges respectively. In order to compute the distillable entanglement after gauging, we note that in the abelian case, upon projecting onto one superselection sector $P_{\vec{r}}$, the gauging map decomposes into a tensor product of isometries $P_{\vec{r}}\cG = \cG^A_{\vec{r}} \otimes \cG^B_{\vec{r}}$ (taking the normalization into account). Hence the depolarized state $\sigma^{\psi_{g,\text{mat}}}$ can be prepared starting from $\sigma^{\psi_{\text{mat}}}$ by LOCC which implies that $E_D(|\psi_{\text{mat}}\rangle) \geq E_D^{\text{gauged}}(|\psi_{g,\text{mat}}\rangle)$. However, again by the fact that with respect to each sector the gauging map is a tensor product of local observables, each local symmetric operator on the matter Hilbert space can be mapped to a local gauge invariant operation~\cite{gaugingpaper} which in turn yields the other inequality $E_D(|\psi_{\text{mat}}\rangle) \leq E_D^{\text{gauged}}(|\psi_{g,\text{mat}}\rangle)$. We find that for abelian theories in the zero-coupling limit the distillable entanglement after gauging is the same as that of the original matter state, if we restrict in this case to symmetric operations (see the Appendix for the detailed argument). 
%
%
%

 For an illustration of how matter degrees can change the value of the physical entanglement content, we consider a $\mathbb{Z}_2$ gauge theory. For a zero hopping term between the matter degrees of freedom, we are effectively in the pure abelian gauge case, with a zero distillable entanglement. Turning on the hopping term induces a positive distillable entanglement. For small interaction strength $\varepsilon$ and large boundary area $\partial A$ we find using perturbation theory (see the Appendix for details):
\begin{align}
  E_D^{\text{gauge}} \approx \partial A\,H(\cos^2(\varepsilon),\sin^2(\varepsilon)) - \log(2)\,.
\end{align}
Hence the physically accessible entanglement obeys an area law featuring the same topological correction as the full entanglement entropy, a fact we expect to hold for arbitrary finite abelian groups and generic translation invariant Hamiltonians at zero gauge coupling (see the Appendix for an argument).


%
\noindent\emph{Conclusions}---%
%
%
%
We addressed the issue of entanglement for lattice gauge theories from an operational standpoint.
As we have shown, it is the entanglement in the multiplicity spaces of the local gauge constraints that can indeed be distilled by an LOCC protocol. Interestingly, the distillable entanglement violates subadditivity. Moreover, we calculated its value for zero-coupling gauge theories -- abelian and nonabelian, with and without matter. We stress that our general results are valid both for discrete lattice gauge spin systems and for relativistic gauge quantum field theories that are regulated by a spatial lattice formulation\footnote{In the latter case for infinite groups $|G|\rightarrow \infty $ one can regularize the local edge Hilbert spaces, either by considering a finite subgroup, or by truncating the allowed irreducible representations $r$.} \cite{PhysRevD.11.395,Kogut:1979jp,BaezLGT}. 

One of the motivations for this letter was the firewall paradox \cite{firewall,braunstein} that crucially hinges on the violation of the monogamy of entanglement on a local Hilbert space. It should be interesting to reformulate the paradox for gauge field theories in light of our results.
%

During the completion of this work, we became aware of the recent contribution~\cite{2015arXiv151007455S}, that has a partial overlap with our results.


\noindent\emph{Acknowledgements}---%
We acknowledge inspiring discussions with David Dudal and Henri Verschelde and thank C.-T.~Ma for helpful comments on an earlier version of this paper. This work was supported by the Austrian Science Fund (FWF) through grants ViCoM and FoQuS and the EC through grants QUTE and SIQS. J.H. and M.M. are supported by the FWO.




\onecolumngrid

\appendix
%

\section{Kogut-Susskind formulation of lattice theories and bipartite cuts}

To set up the Kogut-Susskind (K-S) Hamiltonian formulation of lattice gauge theory~\cite{PhysRevD.11.395,Kogut:1979jp,BaezLGT} for a general compact group $G$, we start from a certain lattice $(V,L)$ consisting of vertices $v \in V$ that are connected by edges $e \in L$. The gauge degrees of freedom are located on the edges: on each edge $e$ we have a local Hilbert space $\mathcal{H}_e \simeq L^2(G)$ spanned by the group basis $\{\ket{g}_e\}_{g\in G}$. We also define a left and right group action for each edge, $L_e(h)\ket{g}_e=\ket{h g}_e$ and $R_e(h)\ket{g}_e=\ket{g h^{-1}}_e$.
Notice that $[L_e(h_1),R_e(h_2)]=0$, while in general both $[L_e(h_1),L_e(h_2)]\neq 0$ and $[R_e(h_1),R_e(h_2)]\neq 0$. The matter degrees of freedom are located on the vertices: on each vertex $v$ the local Hilbert space $\mathcal{H}_m$ consists of a representation space of the group $G$, with a certain number of irreducible representations and the local group action $V_v(h)$ defined accordingly.  One then obtains the Hilbert space $\mathcal{H}$ simply by taking the direct product of all local Hilbert spaces: $\mathcal{H}=\otimes_e \mathcal{H}_e \otimes_v\mathcal{H}_v$.

 The local gauge symmetry is of course what makes a theory a gauge theory. In the K-S Hamiltonian formulation this symmetry is implemented by time-independent local gauge transformations. To define them we have to assign an orientation to the edges (see the figure). The local gauge transformation $U_v(g)$ at a vertex $v$ is then defined as $U_v(g)=V_v(g)\otimes_{e\in E^+_v} R_e(g)\otimes_{e \in E^-_v}L_{e}(g)$, where the sets $E^+_v$ and $E^-_v$ consist respectively of all incoming and outgoing edges of the vertex $v$.  For the simple case of a square lattice that we consider in the figure, $E^+_v$ contains the edges at the lefthand side and at the bottom, while $E^-_v$ consists of the edges at righthand side and the top. Notice also that $[U_{v_1}(g_1), U_{v_2}(g_2)]=0$ for any two different vertices $v_1, v_2$.
Now the physical Hilbert space $\mathcal{H}_{\text{phys}}$ is defined as the subspace of $\mathcal{H}$ consisting of all gauge-invariant states, that is all states $\ket{\psi}$ that obey the gauge constraints:
\begin{align}
  U_v(g)\ket{\psi}=\ket{\psi}\,,\label{eq:constraint}
\end{align}
for all vertices $v$ and all group elements $g$.

Let us now concentrate on a bipartite cut, dividing the full full system into a connected spatial region $A$ and its complement $B$, connected by $n$ boundary edges. We define the boundary vertices $v_b$ as those vertices for which the gauge transformations have non-trivial support on both $A$ and $B$: $U_{v_b}(g)=U^A_{v_b}(g)\otimes U^B_{v_b}(g)$. Notice that $U^A_{v_b}(g)$ and $U^B_{v_b}(g)$  can take different forms: $L_{e_1}(g)$, $L_{e_1}(g)\otimes R_{e_2}(g)$, $L_{e_1}(g)\otimes R_{e_2}(g)\otimes V_{v_b}(g),\ldots$ depending on the vertex $v_b$ and on the specific partition. For simplicity, we will always explicitly consider partitions as in the figure, where each $U^A_{v_b}(g)$ only has non-trivial support on a single edge. But this will not affect the generality of our conclusions. As basis for the Hilbert space $\mathcal{H}_A$ on $A$ we can then take the orthonormal irreducible representation basis of $\otimes_{v_b} G$ on $A$: $\{\ket{\vec{r},\vec{i
 },\alpha
 }_A\}$. Here $\vec{r}=(r_1,r_2,\ldots,r_n) \in \Lambda^n$ and $\vec{i}=(i_1,i_2,\dots, i_n)$, where $\Lambda$ is the set of equivalence classes of irreducible representations of $G$ and the subindex runs over the $n$ different boundary lattice vertices $v_b$. The group action is implemented as $U_1^A(g_1)\ldots U_n^A(g_n)\ket{\vec{r},(i_1,\dots, i_n),\alpha}_A=\sum_{(j_1,\ldots,j_n)}\Gamma^{r_1}_{j_1i_1}(g_1)\ldots \Gamma^{r_n}_{j_n i_n}(g_n)\ket{\vec{r},(j_1,\dots, j_n),\alpha}_A$.
%
The index $\alpha$ labels the multiplicity of the direct product irreducible representation $\vec{r}$. Notice that this multiplicity can be zero (in which case we will of course not have the corresponding irreducible representation $\vec{r}$ in our basis). In a similar way we can take for $\mathcal{H}_B$ the irreducible representation basis of $\otimes_{v_b} G$ on $B$: $\{\ket{\vec{r},\vec{i},\beta}_B\}$, with the difference that we now label the states according to their transformation under the complex conjugate irreps $\Gamma^{r*}_{ij}(g)$. It is clear that our choice of bases for $A$ and $B$ amounts to the decompositions:
\begin{align}
   \mathcal{H}_A = \oplus_{\vec{r}} \mathcal{H}^{\vec{r}}_{Ag} \otimes \mathcal{H}^{\vec{r}}_{Am}\,,\quad\quad \mathcal{H}_B = \oplus_{\vec{r}} \mathcal{H}^{\vec{r}}_{Bg} \otimes \mathcal{H}^{\vec{r}}_{Bm}\,,
\label{eq:decompAB} \end{align}
with $\mathcal{H}^{\vec{r}}_{Ag}, \mathcal{H}^{\vec{r}}_{Bg}$ the representation spaces for the irrep $\vec{r}$ corresponding to the group indices $\vec{i}$ and $\mathcal{H}^{\vec{r}}_{Am}, \mathcal{H}^{\vec{r}}_{Bm}$ the multiplicity spaces on $A$ and $B$.
Using the implications of the Peter-Weyl theorem, in particular the relation
\begin{equation}
  \int_G dg \sqrt{d_r d_r'}\Gamma^{r*}_{ij}(g)\Gamma^{r'}_{lk}(g)=\delta_{rr'}\delta_{il}\delta_{jk}\,,
  \label{eq:Peter-Weyl}
\end{equation}
with $d_r$ the dimension of the irrep $r$ of $G$ and $\int\!dg$ the group-integral \footnote{In our general expressions we explicitly consider continuous groups. To translate to the discrete case, one simply has to replace the Haar-measure integral $\int\!dg$ with the discrete sum $\frac{1}{|G|}\sum_g$; and correspondingly one has to replace the Dirac-delta function in $\Scp{g}{h} =\delta(g-h)$ with a Kronecker-delta $\Scp{g}{h}=\delta_{g,h}$. },
we can now easily show that a complete orthonormal basis for the gauge invariant states on $\mathcal{H}$, i.e. the states $\ket{\psi}$ for which $U^A_{v_b}(g)\otimes U^B_{v_b}(g)\ket{\psi}=\ket{\psi}$ for all boundary vertices $v_b$ and all group elements $g$, consists of $\{\ket{\vec{r},\alpha,\beta}\equiv1/\sqrt{d_{\vec{r}}}\sum_{\vec{i}}\ket{\vec{r},\vec{i},\alpha}_A \ket{\vec{r},\vec{i},\beta}_B\}\,,$ where $d_{\vec{r}}\equiv \prod_n d_{r_n}$. This then amounts to the decomposition:
\begin{align}
  \label{eq:decompphyshilbertspace}
  \mathcal{H}_{\text{phys}}= \bigoplus_{\vec{r} \in \Lambda^n} \left(\ket{\phi^{\vec{r}}}_{g AB}\otimes \mathcal{H}^{\vec{r}}_{Am} \otimes \mathcal{H}^{\vec{r}}_{Bm}\right) \,,
\end{align}
with $\ket{\phi^{\vec{r}}}_{g AB}=\sum_{\vec{i}}\ket{\vec{i}}_A\ket{\vec{i}}_B/\sqrt{d_{\vec{r}}}$, the maximally entangled state in the representation space $\mathcal{H}^{\vec{r}}_{Ag} \otimes \mathcal{H}^{\vec{r}}_{Bg}$.

\section{Distillable entanglement for nonabelian groups}

First we write for the projector $\cQ_{\cH_{\text{phys}}}=P_A P_B \prod_{v_b}P_{v_b}$, with $P_A, P_B$ the product of all vertex projectors, having only non-trivial support on $A$ or $B$. Also, for convenience we now consider the orientations on all boundary edges $e_b$ of $A$ pointing inwards, such that $U_{v_b}^A(g_{v_b})=L(g_{v_b})$ for all boundary gauge transformations. One can then show that tracing out $B$ results in the following density matrix on $A$:
\begin{eqnarray}
  \rho_A&\equiv& \tr_{B}\ket{\psi_0}\bra{\psi_0}\,\nonumber\\
 &=& \prod_{v_b}\int_G\! dg_{v_b}  \prod^{n}_{v_b=1}U_{v_b}^A(g_{v_b}) P_A \ket{e}_A \bra{e}_A P_A\prod^{n}_{v_b=1} U^{A+}_{v_b}(g_{v_b})\\
 &\equiv&\prod_{v_b}\int_G\! dg_{v_b}\ket{\phi(\{g_{v_b}\})}\bra{\phi(\{g_{v_b}\})}\,,
\end{eqnarray}
with $\ket{e}_A\equiv \ket{e}^{\otimes |L_A|}$. Furthermore it can be seen that the states $\ket{\phi(\{g_{v_b}\})}$ are equivalent under a global symmetry $h$: $\ket{\phi(\{g_{v_b}\})}=\ket{\phi(\{g_{v_b h}\})}$ and that the different equivalence classes are orthogonal: $\Scp{\phi(\{\tilde g_{v_b}\})}{\phi(\{g_{v_b}\})}=\int_G\!dh\,\prod_{v_b}\delta(\tilde g_{v_b}g_{v_b}^{-1}-h)$. This implies that the density matrix is flat \cite{quantumdouble1,quantumdouble2}, with non-zero eigenvalues on the $|G|^{n-1}$ independent orthogonal states $\ket{\phi (g_1,\ldots, g_{n-1},e)}$, leading to an entanglement entropy $S_A=(n-1)\log |G|$ on the full Hilbert space. To calculate the distillable entanglement we have to write $\rho_A$ in the irreducible representation basis. From the Peter-Weyl relation \eqref{eq:Peter-Weyl}, one can construct the appropriate set of orthonormal states:
\begin{equation}
  \ket{\vec{r},\vec{i},\alpha}=\prod_{
 {v_b}\in
 t_G}\!\int_G\!dg_{v_b}\sum_{\vec{j}} c^{\alpha}_{\vec{r},\vec{j}}\Gamma^{r_1}_{i_1,j_1}(g_1)\ldots\Gamma^{r_n}_{i_n,j_n}\ket{\phi(\{g_{v_b}\})}\,.
 \label{eq:irreps}
\end{equation}
 Here the multiplicity index $\alpha=1,\ldots, N^1_{\vec{r}}$ labels the different scalars under the right global group action that appear in the fusion of $\vec{r}$: $\sum_{\vec{k}}\Gamma^{r_1*}_{j_1,k_1}(h)\ldots \Gamma^{r_n*}_{j_n,k_n}(h) c^{\alpha}_{\vec{r},\vec{k}}=c^{\alpha}_{\vec{r},\vec{j}}$, for any $h\in G$. Orthonormality of the states \eqref{eq:irreps} then follows if we take the normalization $\sum_{\vec{k}}c^{\alpha*}_{\vec{r},\vec{k}} c^{\beta}_{\vec{r},\vec{k}}=d_{\vec{r}}\delta_{\alpha,\beta}$.  In this basis $\rho_A$ then reads:
\begin{equation}
\rho_A=\frac{1}{|G|^{n-1}}\sum_{\vec{r},\vec{i},\alpha}\ket{\vec{r},\vec{i},\alpha}\bra{\vec{r},\vec{i},\alpha}\,,
\end{equation}
from which the formula~\eqref{eq:nonabelian} can be read off.

\section{The distillable entanglement after gauging an abelian symmetry}

In this section we consider a matter state $|\psi_{\text{mat}}\rangle$ with a global abelian symmetry and derive the distillable entanglement before and after gauging. We consider a square lattice and make a bipartition by dividing it in a simply connected region $A$ and its complement $B$. A general symmetric state can then be written as
\begin{equation*}
|\psi_{\text{mat}}\rangle = \sum_{r\alpha\beta} c_{r\alpha\beta}|r\alpha\rangle_A|r^*\beta\rangle_B\, ,
\end{equation*}
where $r$ is an irrep of the abelian group labeling the charge in region $A$ and the orthonormal basis states transform under the group action as $|r\alpha\rangle_A \rightarrow \Gamma^r(g)|r\alpha\rangle_A$, with $\Gamma^r(g)$ an irrep of $G$. The dual irrep $r^*$ is then defined via $\Gamma^{r^*}(g) = \bar{\Gamma}^r(g)$, for all $g\in G$. The labels $\alpha$ $(\beta)$ are degeneracy labels that encode for example the position of the charges in $A$ ($B$). We can write the projector onto globally symmetric states as $\Pi_{\text{symm}} = \sum_{r} \Pi_r^A \otimes \Pi_{r^*}^B$ with $\Pi_r^{A,B}$ local projectors in regions $A$, $B$ that project onto the sector with total charge $r$.

The reduced density matrix of region $A$ is now easily obtained as
\begin{eqnarray*}
\rho^A & = & \sum_r p_r |r\rangle\langle r|\otimes\left(\frac{1}{p_r}\sum_{\alpha\alpha'}\left(\sum_{\beta}c_{r\alpha\beta}c^*_{r\alpha'\beta}\right)|\alpha\rangle\langle \alpha'| \right)\\
 & \equiv & \sum_r p_r |r\rangle\langle r|\otimes \rho^A_r\, ,
\end{eqnarray*}
where $p_r$ is a normalization such that $\tr(\rho_r^A) = 1$. The entanglement entropy is then
\begin{equation*}
S(\rho^A) = H(p_r) + \sum_r p_r S(\rho^A_r)\, .\label{eq:globalent}
\end{equation*}
The first term in this expression is the Shannon entropy of the probability distribution $p_r$ of measuring charge $r$ in region $A$. Similar reasoning as in the main text shows that if one restricts to operators in $A$ that commute with the global symmetry then only the second term is the entanglement distillable via LOCC, i.e. $E_D^{\text{symm}}(|\psi_{\text{mat}}\rangle) = \sum_r p_r S(\rho^A_r)$.

We now define the gauged state $|\psi_{g,\text{mat}}\rangle$, which is defined in the Hilbert space consisting of the tensor product of local Hilbert spaces on both the vertices and edges, by applying the gauging map of \cite{gaugingpaper} to the globally symmetric matter state: $|\psi_{g,\text{mat}}\rangle = \mathcal{G}|\psi_{\text{mat}}\rangle$. The map $\mathcal{G}$ creates the equal weight superposition of all gauge field configurations (= irrep assignments) on the edges that are consistent with the charges on the vertices. It is an ``isometry'' on the set of globally symmetric states $\mathcal{G}^\dagger \mathcal{G} = \Pi_{\text{symm}}$. Linearity of the gauging map implies
\begin{equation*}
|\psi_{g,\text{mat}}\rangle = \sum_{r\alpha\beta} c_{r\alpha\beta}\mathcal{G}(|r\alpha\rangle|r^*\beta\rangle)\, ,
\end{equation*}
where the product state $|r\alpha\rangle|r^*\beta\rangle$ will be gauged into an entangled state. Let us now denote the number of edges that connect vertices in $A$ to vertices in $B$ with $n$. The $n$-dimensional vector $\vec{r}$ is then defined to represent a particular configuration of irreps on the edges that connect $A$ to $B$. We now use the fact that for abelian groups the gauging map on a bipartition can be written as
\begin{equation}
\mathcal{G} = \frac{1}{|G|^{\frac{n-1}{2}}}\sum_{\vec{r}}\mathcal{G}^A_{\vec{r}}\otimes \mathcal{G}^B_{\vec{r}}\, ,
\end{equation}
where $\mathcal{G}^{A}_{\vec{r}}$ ($\mathcal{G}^{B}_{\vec{r}}$) map states in $A$ or $B$ to an equal weight superposition of all gauge field configurations consistent with the charges on the vertices and subject to the boundary condition that the $n$ edges along the boundary are fixed to the configuration $\vec{r}$. As a consequence of Gauss' law, these maps are ``isometric'' in the sense that $\mathcal{G}^{A\dagger}_{\vec{r}} \mathcal{G}^{A}_{\vec{r}} = \Pi^A_{|\vec{r}|}$ and $\mathcal{G}^{B\dagger}_{\vec{r}} \mathcal{G}^{B}_{\vec{r}} = \Pi^B_{|\vec{r}|^*}$ where we abuse the notation $|\vec{r}|$ to denote the total charge to which the product of irreps $\vec{r}$ will fuse. The normalization in front of the previous decomposition is explained by observing that there exist $|G|^{n-1}$ different configuration $\vec{r}$ that give rise to the same $|\vec{r}|$.

Applying this decomposition of the gauging map onto $\ket{\psi_{\text{mat}}}$ gives rise to
\begin{equation*}
|\psi_{g,\text{mat}}\rangle = \frac{1}{|G|^{\frac{n-1}{2}}}\sum_r \sum_{\vec{r}}\sum_{r\alpha\beta} c_{r\alpha\beta}\,\mathcal{G}^A_{\vec{r}}|r\alpha\rangle\mathcal{G}^B_{\vec{r}}|r^*\beta\rangle .
\end{equation*}
where now $\mathcal{G}^A_{\vec{r}}$ will annihilate the state $\ket{r\alpha}$ if $r\neq |\vec{r}|$. So defining $|\vec{r}\alpha\rangle = \mathcal{G}_{\vec{r}}|r\alpha\rangle$ and henceforth implicitly using $r=|\vec{r}|$, the gauged state is
\begin{equation}
|\psi_{g,\text{mat}}\rangle = \frac{1}{|G|^{\frac{n-1}{2}}}\sum_{\vec{r}}\sum_{\alpha\beta} c_{r\alpha\beta}\,|\vec{r}\alpha\rangle|\vec{r}\beta\rangle.
\end{equation}
and we obtain the reduced density matrix of subsystem $A$ as
\begin{equation}
\rho^A_g = \sum_{\vec{r}} \frac{p_r}{|G|^{n-1}}\left(\frac{1}{p_r}\sum_{\alpha\alpha'\beta} c_{r\alpha\beta}c^*_{r\alpha'\beta}|\vec{r}\alpha\rangle\langle\vec{r}\alpha'|\right)\, ,
\end{equation}
where $p_r$ is the same as in $\rho^A$ (because of the isometric nature of $\mathcal{G}_{\vec{r}}^A$). The entanglement entropy is then given by
\begin{equation}
S(\rho^A_g) = \log|G|(n-1) + H(p_r) + \sum_r p_r S(\rho^A_r)\, .
\end{equation}
While for the distillable entanglement entropy \eqref{eq:mainresult} we find, since $\sum_{\vec{r}}^{|\vec{r}|=r}p_{\vec{r}}=p_r$ and $\rho^A_{\vec{r}}=\rho^A_r$: \begin{equation} E_D^{\text{gauge}}(|\psi_{g,\text{mat}}\rangle) = \sum_{\vec{r}}p_{\vec{r}}S(\rho^A_{\vec{r}}) = \sum_r p_r S(\rho^A_r)\,. \end{equation}

%
\section{Perturbative Calculations}\label{sec:app}
In this section we calculate the lowest order corrections to the entropy of a $\mathbb{Z}_2$ gauge theory with matter if an interaction between the matter degrees of freedom is turned on. A similar calculation can be done for all abelian gauge theories. This calculation is an illustration of the results obtained in the main text and in the previous section, although we will not make explicit use of it. In a perturbative approach that we explain below we calculate the full entanglement entropy of the zero-coupling gauge theory and show that it can indeed be decomposed in a distillable and non-distillable part, consistent with the general expression that was derived above.

We start with the unperturbed Hamiltonian
$$
H=-\sum_{\text{Plaquettes p}} W_p - \sum_{\text{vertices}} X_v
$$
where $W_p=1-Z^{\otimes 4}$ is the projector onto the even subspace of the four gauge qubits around a plaquette. The ground state of this Hamiltonian is given by the equal superposition of all closed loop configurations on the gauge fields and all matter in the product state $\ket{+}$. The local gauge condition is given by the operators $X^{\otimes 4}_g \otimes X_m$ which act on a matter degree and all neighbouring gauge fields.

We now consider a bipartition $A,B$ of the lattice corresponding to a topologically trivial cut. Let there be $n_A$ edges in $A$, $n_B$ in $B$ and $n$ on the boundary. For the toric code, or any abelian gauge theory, the different superselection sectors can be obtained by fixing the elements on every edge on the boundary in one of the eigenvectors $\ket{+},\ket{-}$ of the electric field operator $X$. We can decompose the ground state $\ket{gs}$ as
$$
\ket{gs}=\frac{1}{\sqrt{2^{n-1}}}\bigoplus_{\vec{r}\: \text{even}}  \ket{A,\vec{r}}\ket{\vec{r}}\ket{B,\vec{r}}.
$$
Here, $\ket{\vec{r}}$ is the product state of the edges of the boundary where the value of every edge is specified by $\alpha$ The vectors $\ket{A,\vec{r}},\ket{B,\vec{r}}$ are given by the equal superpositions of all string configurations in $A$ and $B$ respectively that are compatible with the boundary values. All matter is still in the product state $\ket{+}$.

We can perturb the Hamiltonian with a term $V_{\ell}$ on every edge $\ell$ with $V_{\ell}=Z_m\otimes Z_g\otimes Z_m$. Here the $Z_m$ operators work on the two vertices connected to the edge $\ell$ and $Z_g$ acts on the gauge field on this particular edge. The operators $V_{\ell}$ create excitations of the magnetic fields on neighbouring vertices.

We now approximate the entanglement of the ground state $\ket{gs(\varepsilon)}$ of $H+\varepsilon \sum_{\ell}V_{\ell}$.
The results of \cite{hastings2005quasiadiabatic,hastings2010locality,bachmann2012automorphic} imply that the exact ground state $\ket{gs(\varepsilon)}$ of the Hamiltonian $H+\varepsilon \sum_{\ell}V_{\ell}$ can be obtained by evolving $\ket{gs}$ in time,
$$
\ket{gs(\varepsilon)} = \exp\left(-i\int_0^{\varepsilon}D(x)dx\right)\ket{gs}
$$
with $D(x)$ quasi-local in the sense that the strength of the interaction of $D(x)$ decay superpolynomially as a function of the size of its support. We now quickly argue that instead of focusing on the entanglement of $\ket{gs(\varepsilon)}$ we can calculate the entanglement of a simpler state, called $\ket{\tilde{gs}_{AB}}$ in the remainder of this section. For more details we refer the reader to the references. The argument goes as follows. Using the formula for the entanglement rate \cite{van2013entanglement}, we see that to find the entanglement of $\ket{gs(\varepsilon)}$ it suffices to only evolve with terms in $D(x)$ located close to the boundary between $A,B$. We can usually restrict ourselves to operators located a distance $\sim \log(n)$ from the boundary.  Denote this operator by $D_{AB}$. We now have
\begin{eqnarray}
\exp\left(-i\int_0^{\varepsilon}D_{AB}(x)dx\right) &=& \exp\left(-i\varepsilon D_{AB}(0)\right)\left(\exp\left(+i\varepsilon D_{AB}(0)\right)\exp\left(-i\int_0^{\varepsilon}D_{AB}(x)dx\right)\right)\\
&=&\exp\left(-i\varepsilon D_{AB}(0)\right)\exp\left(i\varepsilon \tilde{H}\right)\,,
\end{eqnarray}
with $||\tilde{H}||\approx \varepsilon\max_{s\in [0,\varepsilon]}\|D_{AB}(0)-D_{AB}(s)\|$, which scales as $\epsilon n\log n$. This motivates us to simplify the evolution operator even further, we henceforth focus on the state
$$
\ket{\tilde{gs}_{AB}}=\exp\left(-i\varepsilon D_{AB}(0)\right)\ket{gs},
$$
which from the bound on the entanglement rate \cite{van2013entanglement} implies an error $\propto \varepsilon ||\tilde{H}||=\varepsilon^2 n\log n$ on the calculated entropy.
The operator $D_{AB}(0)$ can easily be calculated using the formulas in \cite{hastings2010locality,bachmann2012automorphic}. If we denote by  $W_{\ell}$ the operator $Z_m\otimes Z_g\otimes Y_m$, very similar to the operator $V_{\ell}$, then $D_{AB}(0) = \sum_{\ell} W_{\ell}$. As we work on a bipartite lattice, we can equally use an operator with the $Z$ always acting on the matter in one of the part and the $Y$ on the other part. This convention has the benefit that all terms $W_{\ell}$ commute.
We now have
$$\ket{\tilde{gs}_{AB}}=
\exp\left(i\varepsilon \sum_{\ell} W_{\ell}\right)\ket{gs} =\prod_{\ell}\left( \cos(\varepsilon)1+i\sin(\varepsilon)W_{\ell}\right)\ket{gs}.
$$
Since $\exp(i\varepsilon \sum_{\ell} W_{\ell})$ is a unitary generated by a local commuting Hamiltonian, we can first apply all the factors acting exclusively in $A$ or $B$. As these do not change the entanglement we can look at the entanglement of the state
$$
\prod_{\ell \in \partial AB}\left( \cos(\varepsilon)1+i\sin(\varepsilon)W_{\ell}\right)\ket{gs}
$$
where the sum now ranges only over $V_{\ell}$ acting across boundary.
As all matter degrees of freedom in $\ket{gs}$ start in a product state of $\ket{+}$ it is easy to see that the application of this last operator on $\ket{gs}$ is equal to \footnote{This fails whenever $W_{\ell}$ for two different $\ell=\ell'$ share a vertex degree of freedom, i.e. at a corner. This would give rise to corner corrections to the entanglement which we henceforth ignore. Alternatively, for a state on a sphere, with $A$ one half of the sphere, or for $A$ a suitably chosen region on a hexagonal lattice, this problem does not occur.}
$$
\prod_{\ell \in \partial AB}\left(\cos(\varepsilon)1+i\sin(\varepsilon)W_{\ell}\right)\ket{gs}
=\prod_{\ell \in \partial AB}\left(\cos(\varepsilon)1+\sin(\varepsilon)V_{\ell}\right)\ket{gs}.
$$

Let us now compute the entropy of the state $\ket{\tilde{gs}_{AB}}$.
We first focus on an even sector $\alpha$. After applying the gates across the boundary we find that the state in this sector is given by
$$
\frac{1}{\sqrt{2^{n-1}}}\sum_{k \:\text{even}} \sum_{m=1}^{{n}\choose{k}}\cos(\varepsilon)^{n-k}\sin(\varepsilon)^k \ket{A_{k,m},\tilde{\vec{r}}_{k,m})}\ket{\vec{r}}\ket{B_{k,m},\tilde{\vec{r}}_{k,m}}.
$$
Here $\ket{A_{k,m},\tilde{\vec{r}}_{k,m}}$ is the state obtained from $\ket{A,\tilde{\vec{r}}_{k,m}}$ by acting with $Z$ on the $k$ vertices neighbouring edges where $\alpha$ differs from $\tilde{\vec{r}}_{k,m}$ and similar for the state $\ket{B_{k,m},\tilde{\vec{r}}_{k,m}}$.
Moreover this expression is immediately a Schmidt decomposition of the state in the sector $\vec{r}$. The normalization of this state is then given by
$$
\frac{1}{2.2^{n-1}}\left((\cos^2(\varepsilon)+\sin^2(\varepsilon))^n+(\cos^2(\varepsilon)-\sin^2(\varepsilon))^n\right)=\frac{1}{2^n}(1+\cos(2\varepsilon)^n).
$$
Similarly for odd sectors we find
$$
\frac{1}{\sqrt{2^{n-1}}}\sum_{k \:\text{odd}} \sum_{m=1}^{{n}\choose{k}}\cos(\varepsilon)^{n-k}\sin(\varepsilon)^k \ket{A_{k,m},\tilde{\vec{r}}_{k,m})}\ket{\vec{r}}\ket{B_{k,m},\tilde{\vec{r}}_{k,m}}
$$
with normalisation $\frac{1}{2^n}(1-\cos(2\varepsilon)^n)$. The sum over all $p_{\alpha}$ still sums to 1, as we applied a unitary. We can see that the non-distillable part of the entropy is now given by
$$\begin{aligned}
E_{\text{non-distillable}}(\ket{\tilde{gs}_{AB}})&=-2^{n-1}\left(\frac{1+\cos^n(2\varepsilon)}{2^n}\right)\log\left(\frac{1+\cos^n(2\varepsilon)}{2^n}\right)
-2^{n-1}\left(\frac{1-\cos^n(2\varepsilon)}{2^n}\right)\log\left(\frac{1-\cos^n(2\varepsilon)}{2^n}\right)\\
&=(n-1)\log(2)+H(p_{\text{even}},p_{\text{odd}})
\end{aligned}$$
with $p_{\text{even}}=\frac{1+\cos^n(2\varepsilon)}{2}, p_{\text{odd}}=\frac{1-\cos^n(2\varepsilon)}{2}$. We clearly recognize the first two terms obtained in the section of the main text that deals with gauging theories of matter.

We now look at the distillable entanglement. We can look at all sectors $\vec{r}$ independently, look at the usual entanglement in a sector and finally average over the sectors. We start with a sector with $\vec{r}$ even. For a given sector the Schmidt values are given by the probabilities of a sequence of $n$ Bernoulli variables $B_i$, with the extra restriction that the sum of all values $1$ is even. The normalization constant for the probabilities is given by $p_{\text{even}}$. We first state a useful combinatorial identity. Taking the derivative of the equality
$$
\sum_{k\:\text{even}}{{n}\choose{k}}p^{n-k}q^k = \frac{(p+q)^n+(p-q)^n}{2}
$$
to $q$ gives
$$
\sum_{k\:\text{even}}{{n}\choose{k}}kp^{n-k}q^{k-1} = \frac{n(p+q)^{n-1}-n(p-q)^{n-1}}{2}.
$$
The distillable entanglement in the even sector $\vec{r}$ is given by
\begin{align*}
E_{D,\vec{r},\text{even}}(\ket{\tilde{gs}_{AB}}) &= -\sum_{k\:\text{even}} {{n}\choose{k}}\frac{\left(\cos^2(\varepsilon)\right)^{n-k}\left(\sin^2(\varepsilon)\right)^k}{p_{\text{even}}}\log\left(\frac{\left(\cos^2(\varepsilon)\right)^{n-k}\left(\sin^2(\varepsilon)\right)^k}{p_{\text{even}}}\right)\\
&=-\frac{\log\left(\cos^2(\varepsilon)\right)}{p_{\text{even}}}\sum_{k\:\text{even}} {{n}\choose{k}}(n-k)\left(\cos^2(\varepsilon)\right)^{n-k}\left(\sin^2(\varepsilon)\right)^k\\
 & \quad
-\frac{\log\left(\sin^2(\varepsilon)\right)}{p_{\text{even}}}\sum_{k\:\text{even}} {{n}\choose{k}}(k)\left(\cos^2(\varepsilon)\right)^{n-k}\left(\sin^2(\varepsilon)\right)^k\\
& \quad +\log\left(p_{\text{even}}\right)\\
&=
-n\cos^2(\varepsilon)\frac{\log\left(\cos^2(\varepsilon)\right)}{p_{\text{even}}}\frac{1+\cos(2\varepsilon)^{n-1}}{2}
-n\sin^2(\varepsilon)\frac{\log\left(\sin^2(\varepsilon)\right)}{p_{\text{even}}}\frac{1-\cos(2\varepsilon)^{n-1}}{2}+\log\left(p_{\text{even}}\right).
\end{align*}
A similar calculation for $\vec{r}$ odd gives
$$
E_{D,\vec{r},\text{odd}}(\ket{\tilde{gs}_{AB}})=
-n\cos^2(\varepsilon)\frac{\log\left(\cos^2(\varepsilon)\right)}{p_{\text{odd}}}\frac{1-\cos(2\varepsilon)^{n-1}}{2}
-n\sin^2(\varepsilon)\frac{\log\left(\sin^2(\varepsilon)\right)}{p_{\text{odd}}}\frac{1+\cos(2\varepsilon)^{n-1}}{2}+\log\left(p_{\text{odd}}\right).
$$
We conclude that the distillable entanglement is given by
$$
E_D(\ket{\tilde{gs}_{AB}}) = p_{\text{even}}S_{\alpha,\text{even}}+p_{\text{odd}}S_{\alpha,\text{odd}}.
$$
For fixed $\varepsilon$ and in the limit $n \rightarrow \infty$ we see, as mentioned in the main text, $p_{\text{even}}$ and $p_{\text{odd}}$ both go to 1/2, even for very small $\varepsilon$. As this is a combinatorial consequence of translation invariance we indeed expect it to be generic. The expression for the distillable entanglement then becomes very simple,
\begin{align*}
E_D(\ket{\tilde{gs}_{AB}}) &\approx nH(\cos^2(\varepsilon),\sin^2(\varepsilon)) - H(p_{\text{even}},p_{\text{odd}}) \\
&\approx nH(\cos^2(\varepsilon),\sin^2(\varepsilon)) - \log(2)
\end{align*}

We make one final remark concerning the topological entropy. If we  look at the total entanglement we see that the correction to the area law, the so called topological entropy, is still $\log(2)$ as the corrections to this term in the non-distillable and distillable parts of the entropy cancel. Notice that part of the topological entropy is accessible in the interacting state. For fixed $\varepsilon$ and $n \rightarrow \infty$ we even find the usual term $-\log(2)$ as a contribution to the distillable entanglement. We therefore see that the distillable entanglement of the ground state of the perturbed Hamiltonian also features the usual topological correction.

We expect that a similar result holds in the case of arbitrary finite abelian groups for generic translation invariant Hamiltonians at zero gauge coupling. The argument goes as follows.  We can construct any zero-coupling gauge theory state by applying the gauging map $\mathcal{G}$ on a particular globally symmetric matter state $|\psi_{\text{mat}}\rangle$. Let us then take a matter state in the trivial phase that we obtain by applying a low depth quantum circuit (respecting the symmetry) on a globally symmetric product state. This state obeys an area law: $S(\rho^A)=c \partial A=c n $, for some constant $c$. Furthermore, for a {\em generic} circuit we expect a maximal Shannon entropy $H(p_r)\approx \log |G|$ for the distribution of the resulting global charge among the different sectors. See the calculation on $\mathbb{Z}_2$ in the Appendix for an explicit illustration of this. We then find for the distillable entanglement of the gauged state:
\begin{equation}
  E_D^{\textrm{gauge}}= S(\rho^A)-H(p_r)=c \partial A - \log|G|\,,
\end{equation}
it is the last term that we can identify as the topological correction.

\bibliographystyle{apsrev4-1}
\bibliography{library}

\end{document}